\documentclass[11pt]{article}
\usepackage{amsmath,amsxtra}
\usepackage{amscd}
\usepackage{graphicx}
\usepackage{latexsym}
\usepackage{amsmath,amsthm,amsfonts,amssymb,cite}%,showlabels}
\usepackage{latexsym}
\usepackage{amsmath}

\theoremstyle{remark}

\newtheorem*{rem}{Remark}
\numberwithin{equation}{section}
\begin{document}
\hoffset = -2.4truecm \voffset = -2truecm
\renewcommand{\baselinestretch}{1.2}
\newcommand{\mb}{\makebox[10cm]{}\\ }
\date{}
%%%%% DOCUMENT SPECIFIC DEFINITIONS

%  Theorems, Lemmas and the like, should be typeset in italic
\newtheorem{theorem}{Theorem}
\newtheorem{proposition}{Proposition}
\newtheorem{lemma}{Lemma}
\newtheorem{definition}{Definition}

%%%%% END DOCUMENT SPECIFIC DEFINITIONS
%\renewcommand{\square}{\hfill$\Box$\vspace{2ex}}
%\renewcommand{\Theta}{\Ta}

\title{Quasideterminant solutions of a non-Abelian Toda lattice and kink solutions of a matrix sine-Gordon equation}
\author{C.X. Li$^{1,2}$ and J.J.C. Nimmo$^{1}$\\
$^{1}$Department of Mathematics, \\
University of Glasgow \\
Glasgow G12 8QW, UK\\
$^{2}$Department of Mathematics,\\
Capital Normal University\\
Beijing 100037, CHINA}
\date{}
 \maketitle
\begin{abstract}
Two families of solutions of a generalized non-Abelian Toda
lattice are considered. These solutions are expressed in terms
of quasideterminants, constructed by means of Darboux and binary
 Darboux transformations. As an example of the application of these
 solutions, we consider the 2-periodic reduction to a matrix
 sine-Gordon equation. In particular, we investigate the interaction properties of polarized kink solutions.
\end{abstract}

\section{Introduction}
There has been great interest in
noncommutative versions of some well-known soliton equations, such
as the KP equation, the KdV equation and the Hirota-Miwa equation
\cite{K,P,S,WW1,WW2,WW3,H,HT1,DH,JN,GN1,GN2}. Often, these noncommutative
versions are obtained simply by removing the assumption that the coefficients in the Lax pair of the
commutative equation commute.

The non-Abelian Toda lattice
\begin{align}
U_{n,x}+U_nV_{n+1}-V_nU_n&=0,\label{TD1}\\
V_{n,t}+U_{n-1}-U_n&=0,\label{TD2}
\end{align}
was first studied in \cite{MK}.  A Darboux transformations for this system was given by \cite{Salle}. In \cite{NW}, the following generalization
\begin{align}
U_{n,x}+U_nV_{n+1}-V_nU_n&=0,\label{ncTD1}\\
V_{n,t}+\alpha_nU_{n-1}-U_n\alpha_{n+1}&=0,\label{ncTD2}
\end{align}
was studied and the Darboux and binary Darboux transformations were obtained. We note that in general, $\alpha_n$ is not a scalar, but is independent of $t$.
In the case that $U_n$, $V_n$ and $\alpha_n$ are scalars, it is easy to show,
by setting $\alpha_{n+1}U_n=e^{-\theta_n}$ and eliminating $V_n$, that \eqref{ncTD1}--\eqref{ncTD2}
becomes the standard two-dimensional Toda lattice equation
\begin{align}
\theta_{n,xt}-e^{-\theta_{n-1}}+2e^{-\theta_n}-e^{-\theta_{n+1}}=0.\label{TD}
\end{align}

Introducing new variables $X_n$ where
\begin{equation}
U_n=X_nX_{n+1}^{-1},\quad V_n=X_{n,x}X_n^{-1},\label{DVT}
\end{equation}
\eqref{ncTD1}--\eqref{ncTD2} can be rewritten as
\begin{equation}
(X_{n,x}X_n^{-1})_t+\alpha_nX_{n-1}X_n^{-1}-X_nX_{n+1}^{-1}\alpha_{n+1}=0.\label{ncTD}
\end{equation}
From now on, we will refer to \eqref{ncTD} as the non-Abelian Toda lattice.
One type of quasideterminant solutions of \eqref{ncTD} were found in \cite{EGR}. We will show how these  (quasiwronskian) solutions arise from the Darboux transformation and consider a second type of quasideterminant, which we call quasigrammian, solutions obtained using the binary Darboux transformation.

It is well known that the $2$-periodic reduction of the standard two dimensional Toda
lattice leads to the scalar sine-Gordon equation. In the
same way, the $2$-periodic reduction of the non-Abelian Toda lattice
\eqref{ncTD} leads to a noncommutative sine-Gordon equation.
This equation has been studied already in a number of papers
\cite{EGR,LO,AVA,GP,GMPT,ZA,HMNP,CCMM} concerning both the matrix and
the Moyal product versions. Here we only consider in detail the matrix version.

Recently, a matrix KdV equation was considered in \cite{VMG}. A multisoliton solution was found by using the inverse scattering method. In particular, the properties of one- and two-soliton solutions expressed in terms of projection matrices were investigated. We will apply some of these ideas to the matrix sine-Gordon equation to study the interaction of its kink solutions.

%Noticing that a similar nonabelian generalization of the Toda
%lattice equation and its Darboux transformation were also presented
%in \cite{MS}.
%\begin{equation*}
%(g_{nt}g_n^{-1})_\zeta+g_{n+1}g_n^{-1}-g_ng_{n-1}^{-1}=0,
%\end{equation*}
%where $g$ is an $M\times M$ invertible matrix function.

The paper is organized as follows. In Section 2, some properties of
quasideterminants used in the paper are described. In Section 3, we
present quasiwronskian solutions to the non-Abelian Toda lattice
constructed by iterating Darboux transformations and in Section 4,
we present quasigrammian solutions to the system by using the
related binary Darboux transformation. In the rest of the paper we
consider the 2-periodic reduction to a matrix sine-Gordon equation.
In particular we consider the matrix kink solutions obtained from
the quasigrammian solutions, we show that kink solutions for the
matrix sine-Gordon equation emerge intact from interaction apart
from change of polarization and phase.

\section{Preliminaries}\label{sec:prelim}
In this short section we recall some of the key elementary
properties of quasideterminants. The reader is referred to the
original papers \cite{GR,EGR,GGRL} for a more detailed and general
treatment.
\subsection{Quasideterminants}
An $n\times n$ matrix $A$ over a ring $\mathcal R$ (noncommutative,
in general) has $n^2$ \emph{quasideterminants} written as
$|A|_{i,j}$ for $i,j=1,\dots, n$, which are also elements of
$\mathcal R$. They are defined recursively by
\begin{align}\label{defn}
    |A|_{i,j}&=a_{i,j}-r_i^j(A^{i,j})^{-1}c_j^i,\quad A^{-1}=(|A|_{j,i}^{-1})_{i,j=1,\dots,n}.
\end{align}
In the above $r_i^j$ represents the $i$th row of $A$ with the $j$th
element removed, $c_j^i$ the $j$th column with the $i$th element
removed and $A^{i,j}$ the submatrix obtained by removing the $i$th
row and the $j$th column from $A$. Quasideterminants can be also
denoted as shown below by boxing the entry about which the expansion
is made
\[
|A|_{i,j}=\begin{vmatrix}
    A^{i,j}&c_j^i\\
    r_i^j&\fbox{$a_{i,j}$}
    \end{vmatrix}.
\]

The case $n=1$ is rather trivial; let $A=(a)$,  say,  and then there
is one quasideterminant $|A|_{1,1}=|\fbox{$a$}|=a$. For $n=2$, let
$A=\begin{pmatrix}   a&b\\c&d\end{pmatrix}$, then there are four
quasideterminants
\begin{align*}
|A|_{1,1}=\begin{vmatrix}
    \fbox{$a$}&b\\
    c&d
    \end{vmatrix}=a-bd^{-1}c,\quad
|A|_{1,2}=\begin{vmatrix}
    a&\fbox{$b$}\\
    c&d
    \end{vmatrix}=b-ac^{-1}d,\\
|A|_{2,1}=\begin{vmatrix}
    a&b\\
    \fbox{$c$}&d
    \end{vmatrix}=c-db^{-1}a,\quad
|A|_{2,2}=\begin{vmatrix}
    a&b\\
    c&\fbox{$d$}
    \end{vmatrix}=d-ca^{-1}b.
\end{align*}
%From this we can obtain the matrix inverse,
%\[
 %   A^{-1}=\begin{pmatrix}
  %    (a-bd^{-1}c)^{-1}&(c-db^{-1}a)^{-1}\\
   %   (b-ac^{-1}d)^{-1}&(d-ca^{-1}b)^{-1}
   % \end{pmatrix},
%\]
%which is then used in the definition of the 9 quasideterminants of a
%$3\times 3$ matrix.
Note that if the entries in $A$ commmute, the
above becomes the familiar formula for the inverse of a $2\times2$
matrix with entries expressed as ratios of determinants. Indeed this
is true for any size of square matrix; if the entries in $A$ commute
then
\begin{equation}\label{commute}
|A|_{i,j}=(-1)^{i+j}\frac{\det(A)}{\det(A^{i,j})}.
\end{equation}

In this paper we will consider only quasideterminants that are
expanded about a term in the last column, most usually the last
entry. For a block matrix
\begin{equation*}\label{block}
 \begin{pmatrix}
      A&B\\
      C&d
    \end{pmatrix}
\end{equation*}
where $d\in\mathcal R$, $A$ is a square matrix over $\mathcal R$ of
arbitrary size and $B$, $C$ are column and row vectors over
$\mathcal R$ of compatible lengths, we have
\begin{equation*}
 \begin{vmatrix}
      A&B\\
      C&\fbox{$d$}
    \end{vmatrix}
    =d-CA^{-1} B.
\end{equation*}

%\subsection{Invariance under row and column operations}
%The quasideterminants of a matrix have invariance properties similar
%to those of determinants under elementary row and column operations
%applied to the matrix. Consider the following quasideterminant of an
%$n\times n $ matrix;
%\begin{equation}\label{invariance}
 %   \begin{vmatrix}
  %  \begin{pmatrix}
   %   E&0\\
    %  F&g
    %\end{pmatrix}
    %\begin{pmatrix}
     % A&B\\
      %C&d
    %\end{pmatrix}
    %\end{vmatrix}_{n,n}=
    %\begin{vmatrix}
     % EA&EB\\
      %FA+gC&FB+gd
    %\end{vmatrix}_{n,n}=g(d-CA^{-1}B)=g
    %\begin{vmatrix}
     % A&B\\
      %C&d
    %\end{vmatrix}_{n,n}.
%\end{equation}
%The above formula can be used to understand the effect on a
%quasideterminant of certain elementary row operations involving
%multiplication on the left. This formula excludes those operations
%which add left-multiples of the row containing the expansion point
%to other rows since there is no simple way to describe the effect of
%these operations. For the allowed operations however, the results
%can be easily described; left-multiplying the row containing the
%expansion point by $g$ has the effect of left-multiplying the
%quasideterminant by $g$ and all other operations leave the
%quasideterminant unchanged. There is analogous invariance under
%column operations involving multiplication on the right.

%and similarly
%\[
%    \begin{vmatrix}
%    \begin{pmatrix}
%      A&B\\
%      C&d
%    \end{pmatrix}
%    \begin{pmatrix}
%      E&F\\
%      0&g
%    \end{pmatrix}
%    \end{vmatrix}_{nn}=
%    \begin{vmatrix}
%      A&B\\
%      C&d
%    \end{vmatrix}_{nn} g
%\]

\subsection{Noncommutative Jacobi Identity}
There is a quasideterminant version of Jacobi's identity for
determinants, called the noncommutative Sylvester's Theorem  by
Gelfand and Retakh \cite{GR}. The simplest version of this identity
is given by
\begin{equation}\label{nc syl}
    \begin{vmatrix}
      A&B&C\\
      D&f&g\\
      E&h&\fbox{$i$}
    \end{vmatrix}=
    \begin{vmatrix}
      A&C\\
      E&\fbox{$i$}
    \end{vmatrix}-
    \begin{vmatrix}
      A&B\\
      E&\fbox{$h$}
    \end{vmatrix}
    \begin{vmatrix}
      A&B\\
      D&\fbox{$f$}
    \end{vmatrix}^{-1}
    \begin{vmatrix}
      A&C\\
      D&\fbox{$g$}
    \end{vmatrix}.
\end{equation}
As a direct result, we have
%\begin{equation}\label{grow}
   % \begin{vmatrix}
    %  A&B&0\\
    %  C&\fbox{$d$}&0\\
    %  E&f&1
   % \end{vmatrix}
   % =\begin{vmatrix}
   %   A&B\\
   %   C&\fbox{$d$}
  %  \end{vmatrix}
%-    \begin{vmatrix}
  %    A&0\\
   %   C&\fbox{0}
  %  \end{vmatrix}
  %  \begin{vmatrix}
   %   A&0\\
  %    E&\fbox{1}
  %  \end{vmatrix}^{-1}
  %  \begin{vmatrix}
   %   A&B\\
   %   E&\fbox{$f$}
   % \end{vmatrix}
%=
   % \begin{vmatrix}
   %   A&B\\
   %   C&\fbox{$d$}
   % \end{vmatrix}
%\end{equation}
the homological relation
\begin{align}
\begin{vmatrix}
      A&B&C\\
      D&f&\fbox{$g$}\\
      E&h&i
    \end{vmatrix}
&=      \begin{vmatrix}
      A&B&0\\
      D&f&\fbox{0}\\
      E&h&1
    \end{vmatrix}
    \begin{vmatrix}
      A&B&C\\
      D&f&g\\
      E&h&\fbox{$i$}
    \end{vmatrix}.\label{HLR}
\end{align}
\subsection{Quasi-Pl\"{u}cker coordinates}
Given an $(n+k)\times n$ matrix $A$, denote the $i$th row of $A$ by
$A_i$, the submatrix of $A$ having rows with indices in a subset $I$
of $\{1,2,\dots,n+k\}$ by $A_I$ and
$A_{\{1,\dots,n+k\}\backslash\{i\}}$ by $A_{\hat\imath}$. Given
$i,j\in\{1,2,\dots,n+k\}$ and $I$ such that $\#I=n-1$ and $j\notin
I$, one defines the \emph{(right) quasi-Pl\"{u}cker coordinates}
\begin{equation}\label{rplucker}
    r^I_{ij}=r^I_{ij}(A):=
    \begin{vmatrix}
    A_I\\
    A_i
    \end{vmatrix}_{ns}
    \begin{vmatrix}
    A_I\\
    A_j
    \end{vmatrix}_{ns}^{-1}=-
    \begin{vmatrix}
    A_I&0\\
    A_i&\fbox{0}\\
    A_j&1
    \end{vmatrix},
\end{equation}
for any column index $s\in\{1,\dots,n\}$. The final equality in
\eqref{rplucker} comes from an identity of the form \eqref{nc syl}
and proves that the definition is independent of the choice of $s$.

%The following are easy consequence of the definition:
%\begin{align}
%\label{rprop 1}
 % r^I_{ij}&=0\quad\text{if $i\in I$, (and is not defined if $j$ were in $I$)}\\
%\label{rprop 2}
 % r^I_{ii}&=1\\
%\label{rprop 3}
 % r^I_{ji}&=\left(r^I_{ij}\right)^{-1}\\
%\label{rprop 4}
 % r^I_{ij}r^I_{jk}&=r^I_{ik}.
%\end{align}

\begin{rem}
A useful consequence of \eqref{rplucker} is the
identity
\begin{equation}\label{inv}
    \begin{vmatrix}
    A^I&0\\
    A^i&\fbox{0}\\
    A^j&1
    \end{vmatrix}^{-1}=
    \begin{vmatrix}
    A^I&0\\
    A^i&1\\
    A^j&\fbox{0}
    \end{vmatrix},
\end{equation}
which shows that quasideterminants of this form may be inverted very
simply.
\end{rem}

\section{Solutions obtained by Darboux transformations}
The non-Abelian Toda lattice \eqref{ncTD1}--\eqref{ncTD2} has Lax pair
\begin{align}
\phi_{n,x}&=V_n\phi_n+\alpha_n\phi_{n-1},\label{LP1}\\
\phi_{n,t}&=U_n\phi_{n+1}.\label{LP2}
\end{align}
Let $\theta_{n,i}, i=1,...,N$ be a particular set of eigenfunctions
of the linear system and introduce the notation
$\Theta_n=(\theta_{n,1},\cdots,\theta_{n,N})$. The Darboux
transformation, determined by particular solution $\theta_n$, for the non-Abelian Toda lattice is
\begin{eqnarray}
&&\widetilde\phi_n=\phi_n-\theta_n\theta_{n+1}^{-1}\phi_{n+1},\\
&&\widetilde V_n=V_n+\alpha_n\theta_{n-1}\theta_{n}^{-1}-\theta_{n}\theta_{n+1}^{-1}\alpha_{n+1},\\
&&\widetilde U_n=U_n-(\theta_{n}\theta_{n+1}^{-1})_t
=\theta_{n}\theta_{n+1}^{-1}U_{n+1}\theta_{n+2}\theta_{n+1}^{-1},\\
&&\widetilde X_n=\theta_{n}\theta_{n+1}^{-1}X_{n+1}.
\end{eqnarray}
This may be iterated by defining
\begin{align}
\phi_n[k+1]&=\phi_n[k]-\theta_n[k]\theta_{n+1}[k]^{-1}\phi_{n+1}[k],\label{DT1}\\
%&&V_n[k+1]=V_n[k]+\alpha_n\theta_{n-1}[k]\theta_{n}[k]^{-1}-\theta_{n}[k]\theta_{n+1}[k]^{-1}\alpha_{n+1},\\
%&&U_n[k+1]=U_n[k]-(\theta_{n}[k]\theta_{n+1}[k]^{-1})_t
%=\theta_{n}[k]\theta_{n+1}[k]^{-1}U_{n+1}[k]\theta_{n+2}[k]\theta_{n+1}[k]^{-1},\\
X_n[k+1]&=\theta_{n}[k]\theta_{n+1}[k]^{-1}X_{n+1}[k],\label{DT2}
\end{align}
where $\phi_n[1]=\phi_n, X_n[1]=X_n$
%V_n[1]=V_n$,
%$U_n[1]=U_n$
and
\begin{equation}
\theta_{n}[k]=\phi_n[k]|_{\phi_n\rightarrow \theta_{n,k}}.
\end{equation}
In particular,
\begin{align}
\phi_n[2]&=\phi_n-\theta_{n,1}\theta_{n+1,1}^{-1}\phi_{n+1},\label{DT10}\\
X_n[2]&=\theta_{n,1}\theta_{n+1,1}^{-1}X_{n+1}.\label{DT20}
\end{align}

%The expressions for the $N$-times repeated Darboux transformation
%take the form
%\begin{eqnarray}
%&&\Phi_n[N]=\Phi_n-\sigma_{n,1}\Phi_{n+1}-...-\sigma_{n,N}\Phi_{n+N},\\
%&&U_n[N]=U_n-(\sigma_{n,1})_t,\\
%&&V_n[N]=V_n+\alpha_n\sigma_{n-1,1}-\sigma_{n,1}\alpha_{n+1},\\
%&&X_n[N]=(-1)^{N-1}\sigma_{n,N}X_{n+N}.
%\end{eqnarray}
%where $\sigma_{n,k}$ are defined by he system of linear algebraic
%equations
%\begin{equation}
%\sigma_{n,1}\theta_{n+1,j}+...+\sigma_{n,N}\theta_{n+N,j}=\theta_{n,j}
%\quad j=1,2,...,N.
%\end{equation}

In what follows, we will show by induction that the results of $N$
repeated Darboux transformation $\phi_n[N+1]$ and $X_n[N+1]$ can be
expressed as in closed form as quasideterminants
\begin{equation}
\phi_n[N+1]=
\begin{vmatrix}
\Theta_{n}&\fbox{$\phi_n$}\\
\Theta_{n+1}&\phi_{n+1}\\
\vdots&\vdots\\
\Theta_{n+N}&\phi_{n+N}
\end{vmatrix},\quad
X_n[N+1]=(-1)^{N}
\begin{vmatrix}
\Theta_n&\fbox{$0$}\\
\Theta_{n+1}&0\\
\vdots&\vdots\\
\Theta_{n+N}&1
\end{vmatrix}X_{n+N}.\label{QW}
\end{equation}
The initial case $N=1$ follows directly from \eqref{DT10}--\eqref{DT20}. Also using the noncommutative Jacobi identity \eqref{nc syl} and the homological relation \eqref{HLR} we have
\begin{align*}
\phi_n[N+2]&=\phi_n[N+1]-\theta_n[N+1]\theta_{n+1}[N+1]^{-1}\phi_{n+1}[N+1]\\
&=\begin{vmatrix}
\Theta_{n}&\fbox{$\phi_n$}\\
\Theta_{n+1}&\phi_{n+1}\\
\vdots&\vdots\\
\Theta_{n+N}&\phi_{n+N}
\end{vmatrix}-\begin{vmatrix}
\Theta_{n}&\fbox{$\theta_{n,N+1}$}\\
\Theta_{n+1}&\theta_{n+1,N+1}\\
\vdots&\vdots\\
\Theta_{n+N}&\theta_{n+N,N+1}
\end{vmatrix}\begin{vmatrix}
\Theta_{n+1}&\fbox{$\theta_{n+1,N+1}$}\\
\Theta_{n+2}&\theta_{n+2,N+1}\\
\vdots&\vdots\\
\Theta_{n+N+1}&\theta_{n+N+1,N+1}
\end{vmatrix}^{-1}\begin{vmatrix}
\Theta_{n+1}&\fbox{$\phi_{n+1}$}\\
\Theta_{n+2}&\phi_{n+2}\\
\vdots&\vdots\\
\Theta_{n+N+1}&\phi_{n+N+1}
\end{vmatrix}\\
&=\begin{vmatrix}
\Theta_{n}&\fbox{$\phi_n$}\\
\Theta_{n+1}&\phi_{n+1}\\
\vdots&\vdots\\
\Theta_{n+N}&\phi_{n+N}
\end{vmatrix}-\begin{vmatrix}
\Theta_{n}&\fbox{$\theta_{n,N+1}$}\\
\Theta_{n+1}&\theta_{n+1,N+1}\\
\vdots&\vdots\\
\Theta_{n+N}&\theta_{n+N,N+1}
\end{vmatrix}
\begin{vmatrix}
\Theta_{n+1}&\theta_{n+1,N+1}\\
\Theta_{n+2}&\theta_{n+2,N+1}\\
\vdots&\vdots\\
\Theta_{n+N+1}&\fbox{$\theta_{n+N+1,N+1}$}
\end{vmatrix}^{-1}\begin{vmatrix}
\Theta_{n+1}&\phi_{n+1}\\
\Theta_{n+2}&\phi_{n+2}\\
\vdots&\vdots\\
\Theta_{n+N+1}&\fbox{$\phi_{n+N+1}$}
\end{vmatrix}\\
&=\begin{vmatrix}
\Theta_{n}&\theta_{n,N+1}&\fbox{$\phi_n$}\\
\Theta_{n+1}&\theta_{n+1,N+1}&\phi_{n+1}\\
\vdots&\vdots\\
\Theta_{n+N+1}&\theta_{n+N+1,N+1}&\phi_{n+N+1}
\end{vmatrix}
\end{align*}
and
\begin{align*}
X_n[N+2]&=\theta_{n}[N+1]\theta_{n+1}[N+1]^{-1}X_{n+1}[N+1]\\
&=(-1)^N\begin{vmatrix}
\Theta_{n}&\fbox{$\theta_{n,N+1}$}\\
\Theta_{n+1}&\theta_{n+1,N+1}\\
\vdots&\vdots\\
\Theta_{n+N}&\theta_{n+N,N+1}
\end{vmatrix}\begin{vmatrix}
\Theta_{n+1}&\fbox{$\theta_{n+1,N+1}$}\\
\Theta_{n+2}&\theta_{n+2,N+1}\\
\vdots&\vdots\\
\Theta_{n+N+1}&\theta_{n+N+1,N+1}
\end{vmatrix}^{-1}\begin{vmatrix}
\Theta_{n+1}&\fbox{$0$}\\
\Theta_{n+2}&0\\
\vdots&\vdots\\
\Theta_{n+N+1}&1
\end{vmatrix}X_{n+N+1}\\
&=(-1)^N\begin{vmatrix}
\Theta_{n}&\fbox{$\theta_{n,N+1}$}\\
\Theta_{n+1}&\theta_{n+1,N+1}\\
\vdots&\vdots\\
\Theta_{n+N}&\theta_{n+N,N+1}
\end{vmatrix}\begin{vmatrix}
\Theta_{n+1}&\theta_{n+1,N+1}\\
\vdots&\vdots\\
\Theta_{n+N}&\theta_{n+N,N+1}\\
\Theta_{n+N+1}&\fbox{$\theta_{n+N+1,N+1}$}
\end{vmatrix}^{-1}X_{n+N+1}\\
\intertext{and then using the quasi-Pl\"ucker coordinate formula \eqref{rplucker}, we get}
&=(-1)^{N+1}\begin{vmatrix}
\Theta_{n}&\theta_{n,N+1}&\fbox{$0$}\\
\Theta_{n+1}&\theta_{n+1,N+1}&0\\
\vdots&\vdots\\
\Theta_{n+N+1}&\theta_{n+N+1,N+1}&1
\end{vmatrix}X_{n+N+1}.
\end{align*}
This proves the inductive step and the proof is complete.

\section{Solutions obtained by binary Darboux transformation}
The linear equations \eqref{LP1} and
\eqref{LP2} have the formal adjoints
\begin{align}
-\psi_{n,x}&=V_n^\dagger\psi_n+\alpha_{n+1}^\dagger\psi_{n+1},\label{ALP1}\\
-\psi_{n,t}&=U_{n-1}^\dagger\psi_{n-1}.\label{ALP2}
\end{align}

Following the standard construction of a binary Darboux
transformation, one introduces a potential
$\Omega_n=\Omega(\phi_n,\psi_n)$ satisfying the three conditions
\begin{align}
\Omega(\phi_n,\psi_n)_x&=\psi_{n+1}^\dagger\alpha_{n+1}\phi_n,\label{NP1}\\
\Omega(\phi_n,\psi_n)_t&=-\psi_n^\dagger U_n\phi_{n+1},\label{NP2}\\
\Omega_{n}-\Omega_{n-1}&=-\psi_n^\dagger\phi_n.\label{NP3}
\end{align}
A binary Darboux transformation is then defined by
\begin{align}
\phi_n[N+1]&=\phi_n[N]-\theta_n[N]\Omega(\theta_n[N],\rho_n[N])^{-1}\Omega(\phi_n[N],\rho_n[N]),\label{BDT1}\\
\psi_n[N+1]&=\psi_n[N]-\rho_n[N]\Omega(\theta_{n-1}[N],\rho_{n-1}[N])^{-\dagger}\Omega(\theta_{n-1}[N],\psi_{n-1}[N])^\dagger,\label{BDT2}\\
X_n[N+1]&=(I+\theta_n[N]\Omega(\theta_n[N],\rho_n[N])^{-1}\rho_n[N]^\dagger)X_n[N],\label{BDT3}
\end{align}
where $\phi_n[1]=\phi_n, \psi_n[1]=\psi_n$, $X_n[1]=X_n$ and
\begin{equation}
\theta_n[N]=\phi_n[N]|_{\phi_n\rightarrow\theta_{n,N}}, \quad
\rho_n[N]=\psi_n[N]|_{\psi_n\rightarrow\rho_{n,N}}
\end{equation}
Using the notation $\Theta_n=(\theta_{n,1},\dots,\theta_{n,N})$ and
$P_n=(\rho_{n,1},\dots,\rho_{n,N})$, it is easy to prove by induction
that for $N\ge 1$,
\begin{align}
\phi_n[N+1]&=\begin{vmatrix}
\Omega(\Theta_n,P_n)&\Omega(\phi_n,P_n)\\
\Theta_n&\fbox{$\phi_n$} \end{vmatrix},\label{NE}\\
\psi_n[N+1]&=\begin{vmatrix}
\Omega(\Theta_{n-1},P_{n-1})^\dagger&\Omega(\Theta_{n-1},\psi_{n-1})^\dagger\\
P_n&\fbox{$\psi_n$}
\end{vmatrix}\label{NAE}
\end{align}
and
\begin{equation} \Omega(\phi_n[N+1],\psi_n[N+1])=
\begin{vmatrix}
\Omega(\Theta_n,P_n)&\Omega(\phi_n,P_n)\\
\Omega(\Theta_n,\psi_n)&\fbox{$\Omega(\phi_n,\psi_n)$}
\end{vmatrix}.\label{NP}
\end{equation}
We may thus after $N$ binary Darboux transformations we obtain
\begin{equation}
X_n[N+1]=-\begin{vmatrix}
\Omega(\Theta_n,P_n)&P_n^\dagger\\
\Theta_n&\fbox{$-I$} \end{vmatrix}X_n.\label{QG}
\end{equation}

In fact, we can prove the above results by induction.
\begin{align*}
X_n[N+2]&=(I+\Theta_n[N+1]\Omega(\Theta_n[N+1],P_n[N+1])^{-1}P_n[N+1]^\dagger)X_n[N+1]\\
&=-\left(I+\begin{vmatrix}
\Omega(\Theta_n,P_n)&\Omega(\theta_{n,N+1},P_n)\\
\Theta_n&\fbox{$\theta_{n,N+1}$} \end{vmatrix}
\begin{vmatrix}
\Omega(\Theta_n,P_n)&\Omega(\theta_{n,N+1},P_n)\\
\Omega(\Theta_n,\rho_{n,N+1})&\fbox{$\Omega(\theta_{n,N+1},\rho_{n,N+1})$}
\end{vmatrix}^{-1}\right.\\
&\quad\left.\begin{vmatrix}
\Omega(\Theta_{n-1},P_{n-1})&P_n^\dagger\\
\Omega(\Theta_{n-1},\rho_{n-1,N+1})&\fbox{$\rho_{n,N+1}^\dagger$}
\end{vmatrix}\right)
\begin{vmatrix}
\Omega(\Theta_n,P_n)&P_n^\dagger\\
\Theta_n&\fbox{$-I$}
\end{vmatrix}X_n,
\end{align*}
Noticing
\begin{align*}
&\begin{vmatrix}
\Omega(\Theta_{n-1},P_{n-1})&P_n^\dagger\\
\Omega(\Theta_{n-1},\rho_{n-1,N+1})&\fbox{$\rho_{n,N+1}^\dagger$}
\end{vmatrix}
\begin{vmatrix}
\Omega(\Theta_n,P_n)&P_n^\dagger\\
\Theta_n&\fbox{$-I$}
\end{vmatrix}\\
&=-(\rho_{n,N+1}^\dagger-\Omega(\Theta_{n-1},\rho_{n-1,N+1})\Omega(\Theta_{n-1},P_{n-1})^{-1}P_n^\dagger)
(I+\Theta_n\Omega(\Theta_n,P_n)^{-1}P_n^\dagger)\\
&=-\rho_{n,N+1}^\dagger+\Omega(\Theta_{n-1},\rho_{n-1,N+1})\Omega(\Theta_{n-1},P_{n-1})^{-1}P_n^\dagger\\
&\quad+(\Omega(\Theta_{n},\rho_{n,N+1})-\Omega(\Theta_{n-1},\rho_{n-1,N+1}))\Omega(\Theta_n,P_n)^{-1}P_n^\dagger\\
&\quad+\Omega(\Theta_{n-1},\rho_{n-1,N+1})\Omega(\Theta_{n-1},P_{n-1})^{-1}
(\Omega(\Theta_{n-1},P_{n-1})-\Omega(\Theta_n,P_n))\Omega(\Theta_n,P_n)^{-1}P_n^\dagger\\
&=-\rho_{n,N+1}^\dagger+\Omega(\Theta_{n},\rho_{n,N+1})\Omega(\Theta_n,P_n)^{-1}P_n^\dagger\\
&=-\begin{vmatrix}
\Omega(\Theta_n,P_n)&P_n^\dagger\\
\Omega(\Theta_{n},\rho_{n,N+1})&\fbox{$\rho_{n,N+1}^\dagger$}
\end{vmatrix},
\end{align*}
we have
\begin{equation*}
X_n[N+2]=-\begin{vmatrix}
\Omega(\Theta_n,P_n)&\Omega(\theta_{n,N+1},P_n)&P_n^\dagger\\
\Omega(\Theta_n,\rho_{n,N+1})&\Omega(\theta_{n,N+1},\rho_{n,N+1})&\rho_{n,N+1}^\dagger\\
\Theta_n&\theta_{n,N+1}&\fbox{$-I$}
\end{vmatrix}X_n.
%\left(\begin{vmatrix}
%\Omega(\Theta_n,P_n)&P_n^\dagger\\
%\Theta_n&\fbox{$-I$}
%\end{vmatrix}-
%\begin{vmatrix}
%\Omega(\Theta_n,P_n)&\Omega(\theta_{n,N+1},P_n)\\
%\Theta_n&\fbox{$\theta_{n,N+1}$} \end{vmatrix}
%\begin{vmatrix}
%\Omega(\Theta_n,P_n)&\Omega(\theta_{n,N+1},P_n)\\
%\Omega(\Theta_n,\rho_{n,N+1})&\fbox{$\Omega(\theta_{n,N+1},\rho_{n,N+1})$}
%\end{vmatrix}^{-1}\begin{vmatrix}
%\Omega(\Theta_n,P_n)&P_n^\dagger\\
%\Omega(\Theta_{n},\rho_{n,N+1})&\fbox{$\rho_{n,N+1}$}
%\end{vmatrix}
%\right)
\end{equation*}

\section{Matrix sine-Gordon equation and its kink solutions}

It is well known in the commutative case that one may obtain
reductions by imposing periodic conditions on the $\theta_n$.
Similarly in non-Abelian case, one can make periodic reductions of
\eqref{ncTD}. From now on, we only consider the case that $X_n$ is a
$d\times d$ matrix and $\alpha_n=I_{d\times d}$ and so \eqref{ncTD}
is
\begin{equation}
(X_{n,x}X_n^{-1})_t+X_{n-1}X_n^{-1}-X_nX_{n+1}^{-1}=0.\label{ncTDI}
\end{equation}

The simplest such reduction has period $2$, that is, we take $X_n=X_{n+2}$ and \eqref{ncTDI} gives the system
\begin{align}
(X_{0,x}X_0^{-1})_t+X_{1}X_0^{-1}-X_0X_{1}^{-1}&=0,\label{sG1}\\
(X_{1,x}X_1^{-1})_t+X_{0}X_1^{-1}-X_1X_{0}^{-1}&=0.\label{sG2}
\end{align}
We call this a non-Abelian sinh-Gordon equation since in the
commutative case, it will be seen that $X_0=X_1^{-1}=F_{1}/F_0$ and then $\theta=2\log(F_1/F_0)$ satisfies the standard sinh-Gordon equation
$$
    \theta_{xt}=4\sinh\theta.
$$
By changing $\theta\to i\theta$, we can also obtain the sine-Gordon equation
$$
\theta_{xt}=4\sin\theta.
$$

In what follows, we will construct solutions to \eqref{sG1}--\eqref{sG2} by
reduction of the solutions \eqref{QG} of the non-Abelian Toda
lattice \eqref{ncTDI}. It is clear that \eqref{ncTDI} has vacuum solution $X_n=I$ and \eqref{QG} gives the quasigrammian solutions
\begin{equation} X_n=-\begin{vmatrix}
\Omega(\Theta_n,P_n)&P_n^T\\
\Theta_n&\fbox{$-I$} \end{vmatrix},\label{QGS}
\end{equation}
where $\theta_{n,i}$ and $\rho_{n,i}$ satisfy
\begin{align}
(\theta_{n})_x=\theta_{n-1},\quad (\theta_{n})_t=\theta_{n+1},\quad
(\rho_{n})_x=-\rho_{n+1},\quad (\rho_{n})_t=-\rho_{n-1},\label{linvac}
\end{align}
and $\Omega$ is defined by \eqref{NP1}--\eqref{NP3}.
We choose the simplest non-trivial solutions of \eqref{linvac}
\begin{align*}
\theta_{n,j}&=B_jq_j^{-n}e^{q_jx+{1\over q_j}t} ,\quad
\rho_{n,i}=A_ip_i^ne^{-p_ix-{1\over p_i}t}
\end{align*}
where $A_i$ and $B_j$ are $d\times d$ matrices and then we obtain
\begin{align*}
\Omega(\theta_{n,j},\rho_{n,i})=\delta_{i,j}I+{A_i^tB_jp_i\over
q_j-p_i}\left({p_i\over q_j}\right)^ne^{(q_j-p_i)x+({1\over
q_j}-{1\over p_i})t}.
\end{align*}
The choice of constant of integration as $\delta_{i,j}I$ is needed to effect the periodic reduction we will shortly make. This can also be written as
\begin{align*}
\Omega(\theta_{n,j},\rho_{n,i})=\left({p_i\over
q_j}\right)^n\left(\delta_{i,j}I\left({q_j\over
p_i}\right)^n+{A_i^tB_jp_i\over q_j-p_i}e^{(q_j-p_i)x+({1\over
q_j}-{1\over p_i})t}\right).
\end{align*}

Now using the invariance of a quasideterminant to scaling of its
rows and columns (see e.g. \cite{GGRL}), we get
\begin{align*} X_n
% &=-
% \begin{vmatrix}
% I+{A_1^tB_1p_1\over q_1-p_1}\left({p_1\over
% q_1}\right)^ne^{(q_1-p_1)x+({1\over q_1}-{1\over p_1})t}
%  &    \cdots&     {A_1^tB_Np_1\over q_N-p_1}\left({p_1\over
% q_N}\right)^n e^{(q_N-p_1)x+({1\over q_N}-{1\over p_1})t}&A_1^tp_1^ne^{-p_1x-{1\over p_1}t}\\
% \vdots&\ddots&\vdots&\vdots\\
% {A_N^tB_1p_N\over q_1-p_N}\left({p_N\over
% q_1}\right)^ne^{(q_1-p_N)x+({1\over q_1}-{1\over p_N})t}
%  &    \cdots&   I+{A_N^tB_Np_N\over q_N-p_N}\left({p_N\over
% q_N}\right)^n e^{(q_N-p_N)x+({1\over q_N}-{1\over p_N})t}&A_N^tp_N^ne^{-p_Nx-{1\over p_N}t}\\
% B_1q_1^{-n}e^{q_1x+{1\over q_1}t}&\cdots&B_Nq_N^{-n}e^{q_Nx+{1\over
% q_N}t}&\fbox{$-I$} \end{vmatrix}\\%
% &=-\begin{vmatrix} ({q_1\over p_1})^nI+{A_1^tB_1p_1\over
% q_1-p_1}e^{(q_1-p_1)x+({1\over q_1}-{1\over p_1})t}
%  &    \cdots&     {A_1^tB_Np_1\over q_N-p_1} e^{(q_N-p_1)x+({1\over q_N}-{1\over p_1})t}&A_1^te^{-p_1x-{1\over p_1}t}\\
% \vdots&\ddots&\vdots&\vdots\\
% {A_N^tB_1p_N\over q_1-p_N}e^{(q_1-p_N)x+({1\over q_1}-{1\over
% p_N})t}
%  &    \cdots&   ({q_N\over p_N})^nI+{A_N^tB_Np_N\over q_N-p_N} e^{(q_N-p_N)x+({1\over q_N}-{1\over p_N})t}&A_N^te^{-p_Nx-{1\over p_N}t}\\
% B_1e^{q_1x+{1\over q_1}t}&\cdots&B_Ne^{q_Nx+{1\over
% q_N}t}&\fbox{$-I$} \end{vmatrix}
&=-\begin{vmatrix} \left(\delta_{i,j}\left({q_j\over
p_i}\right)^nI+{A_i^tB_jp_i\over q_j-p_i}e^{(q_j-p_i)x+({1\over
q_j}-{1\over p_i})t}\right)&(A_i e^{-p_ix-{1\over p_i}t})^T\\
(B_je^{q_jx+{1\over q_j}t})&\fbox{$-I$}
\end{vmatrix}
\end{align*}
It is obvious from this expression for $X_n$ that it is $2$ periodic
when $({q_1\over p_1})^2=\cdots=({q_N\over p_N})^2=1$, that is,
$p_i=-q_i=\lambda_i$ for $i=1,\cdots,N$. Therefore, the non-Abelian
sinh-Gordon equation has the solutions
\begin{align}
X_0&=-\begin{vmatrix}
(\delta_{i,j}I-{A_i^tB_j\lambda_i\over(\lambda_i+\lambda_j)
}e^{-(\lambda_i+\lambda_j)x-({1\over \lambda_i}+{1\over
\lambda_j})t})
&(A_i e^{-\lambda_ix-{1\over \lambda_i}t})^T\\
(B_je^{-\lambda_jx-{1\over \lambda_j}t})&\fbox{$-I$}
\end{vmatrix},\label{X0}
\\
X_1&=-\begin{vmatrix}
(-\delta_{i,j}I-{A_i^tB_j\lambda_i\over(\lambda_i+\lambda_j)
}e^{-(\lambda_i+\lambda_j)x-({1\over \lambda_i}+{1\over
\lambda_j})t})
&(A_i e^{-\lambda_ix-{1\over \lambda_i}t})^T\\
(B_je^{-\lambda_jx-{1\over \lambda_j}t})&\fbox{$-I$}
\end{vmatrix}.\label{X1}
\end{align}
From now on we will assume that $A_i=I$ are real and $B_j=ir_jP_j$, where $r_j$ are real scalars, are pure imaginary matrices. In this case, it follows  that $X_0$ and $X_1$ are complex conjugate to one another. For this reason we introduce
\begin{equation}\label{X}
    X=X_0=\bar{X_1}=-\begin{vmatrix}
    (\delta_{i,j}I-{B_j\lambda_i\over(\lambda_i+\lambda_j)
    }e^{-(\lambda_i+\lambda_j)x-({1\over \lambda_i}+{1\over
    \lambda_j})t})
    &(e^{-\lambda_ix-{1\over \lambda_i}t}I)^T\\
    (B_je^{-\lambda_jx-{1\over \lambda_j}t})&\fbox{$-I$}
    \end{vmatrix}.
\end{equation}

Next, we will derive matrix kink solutions for the matrix sine-Gordon
equation using the method applied to study the soliton solutions of the matrix KdV equation in \cite{VMG}.
To get a visual representation of the solution we will consider the matrix $W(x,t)$ defined by
\begin{align}\label{soliton}
iW_x=\bar X_{x}\bar X^{-1}-X_{x}X^{-1}.
\end{align}
We choose this dependent variable so that in the scalar case $W=\theta$, the solution of the sine-Gordon equation.

For $N=1$, \eqref{X} gives
\begin{align}
X&=I+B\left(I-{B\over 2}e^{-2\lambda x-{2\over
\lambda}t}\right)^{-1}e^{-2\lambda x-{2\over \lambda}t}.
\end{align}
We first assume further that $P$ is a projection
matrix (i.e.\ satisfies $P^2=P$). This choice allows us to calculate the inverse matrices in the above expression explicitly using the formula
\begin{align}\label{PR}
(I-a P)^{-1}=I+{a P\over 1-a},
\end{align}
where $a\ne1$ is a scalar and $P$ is any projection matrix.

In this way we find that
\begin{align*}
X=I+{irP\over e^{2\lambda x+{2\over \lambda}t}-ir/2},
\end{align*}
and hence
\begin{align*}
W_x={4\lambda P\over \cosh(2\lambda( x+t/\lambda^2-\phi))}
\end{align*}
where $\phi=\log(r/2)/2\lambda$. Note also that $X\bar X=I$.

Taking one final step, we integrate to obtain the one-kink solution to the matrix sine-Gordon equation
\begin{align}
W=4P\arctan(e^{2\lambda( x+t/\lambda^2-\phi)}).\label{KS1}
\end{align}
\begin{rem}
For the one-kink solution \eqref{KS1}, we call the projection matrix
$P$ its polarization and $\phi$ its phase. In the scalar case, if we
choose $P=1$, \eqref{KS1} is simply the one-kink solution to the
standard sine-Gordon equation.
\end{rem}

For $N=2$, expanding $X$ by the definition \eqref{defn}, we can rewrite $X$ as
\begin{align*}
%&=I+(\begin{array}{cc}
%B_1e^{-\lambda_1x-{1\over\lambda_1}t}&
%B_2e^{-\lambda_2x-{1\over\lambda_2}t}\end{array})
%\left(\begin{array}{cc} I-{B_1\over 2}e^{-2\lambda_1x-{2\over
%\lambda_1}t} &-{B_2\lambda_1\over \lambda_2+\lambda_1}
%e^{-(\lambda_2+\lambda_1)x-({1\over \lambda_2}+{1\over
%\lambda_1})t}\\ -{B_1\lambda_2\over
%\lambda_1+\lambda_2}e^{-(\lambda_1+\lambda_2)x-({1\over
%\lambda_1}+{1\over \lambda_2})t} &   I-{B_2\over 2}
%e^{-2\lambda_2x-{2\over \lambda_2}t}\end{array}\right)^{-1}
%\left(\begin{array}{c} e^{-\lambda_1x-{1\over \lambda_1}t}I\\
%e^{-\lambda_2x-{1\over \lambda_2}t}I
%\end{array}
%\right)\\
X&=I+(B_1e^{-\lambda_1x-{1\over\lambda_1}t},B_2e^{-\lambda_2x-{1\over\lambda_2}t})
\left(
\delta_{i,j}I-{B_j\lambda_i\over(\lambda_i+\lambda_j)
}e^{-(\lambda_i+\lambda_j)x-({1\over \lambda_i}+{1\over
\lambda_j})t}
\right)^{-1}_{2\times 2}
%\left(\begin{array}{cc}
%I-{B_1\over 2
%}e^{-2\lambda_1x-2{1\over \lambda_1}t}&-{B_2\lambda_1\over(\lambda_1+\lambda_2)
%}e^{-(\lambda_1+\lambda_2)x-({1\over \lambda_1}+{1\over
%\lambda_2})t}\\
%-{B_1\lambda_2\over(\lambda_2+\lambda_1)
%}e^{-(\lambda_2+\lambda_1)x-({1\over \lambda_2}+{1\over
%\lambda_1})t}&I-{B_2\over 2
%}e^{-2\lambda_2x-2{1\over \lambda_2}t}
%\end{array}
%\right)^{-1}
\left(\begin{array}{c}
e^{-\lambda_1x-{1\over\lambda_1}t}I\\
e^{-\lambda_2x-{1\over\lambda_2}t}I
\end{array}\right)\\
&=I+(L_1e^{\lambda_1x+{1\over\lambda_1}t},L_2e^{\lambda_2x+{1\over\lambda_2}t})
\left(\begin{array}{c}
e^{-\lambda_1x-{1\over\lambda_1}t}I\\
e^{-\lambda_2x-{1\over\lambda_2}t}I
\end{array}\right)\\
% &=I+K_1e^{-\lambda_1x-{1\over
% \lambda_1}t}+K_2e^{-\lambda_2x-{1\over
% \lambda_2}t}\\
&=I+L_1+L_2,
\end{align*}
and hence
\begin{align*}
L_1(e^{2\lambda_1x+{2\over
\lambda_1}t}I-{1\over 2}B_1)-{\lambda_2\over \lambda_1+\lambda_2}L_2B_1&=B_1,\\
L_2(e^{2\lambda_2x+{2\over \lambda_2}t}I-{1\over
2}B_2)-{\lambda_1\over\lambda_1+\lambda_2}L_1B_2&=B_2.
\end{align*}
In the expressions $B_j=ir_jP_j$, $j=1,2$ we assume that $P_j$ are the rank-1 projection matrices
$$P_j={p_j\otimes q_j\over (p_j,q_j)}={p_jq_j^T\over p_j^Tq_j}$$
and the $d$-vectors $p_j$ and $q_j$ satisfy the condition
$(p_j,q_j)\neq 0$, we can solve for $L_1$ and $L_2$ by using \eqref{PR} to obtain
\begin{align*}
L_1&={(\lambda_1+\lambda_2)\over g}(\lambda_2B_2+(\lambda_1+\lambda_2)g_2I)B_1,\\
L_2&={(\lambda_1+\lambda_2)\over
g}(\lambda_1B_1+(\lambda_1+\lambda_2)g_1I)B_2,
\end{align*}
where
\begin{align*}
g&=(\lambda_1+\lambda_2)^2g_1g_2+\lambda_1\lambda_2r_1r_2\alpha,\quad\text{where}\ \alpha={(p_1,q_2)(p_2,q_1)\over (p_1,q_1)(p_2,q_2)}
\end{align*}
and
\begin{align*}
g_j&=e^{2\lambda_j\theta_j}-{ir_j\over 2}\quad \mbox{for}\quad j=1,2,
\end{align*}
where
\[
\theta_j=x+{1\over \lambda_j^2}t.
\]
Therefore
\begin{align}\label{XI}
X=I+{(\lambda_1+\lambda_2)\over
g}(\lambda_1B_1B_2+\lambda_2B_2B_1+(\lambda_1+\lambda_2)(g_1B_2+g_2B_1)).
\end{align}

We now investigate the behaviour of $X$ as $t\rightarrow \pm\infty$. We will use the fact that $W$ is invariant under the transformation $X\to XC$ for any constant matrix $C$ and assume, without loss of generality, that $0<\lambda_1< \lambda_2$. In the calculations that follow, we will demonstrate that kinks emerge from the interaction and undergo phase-shifts as in the scalar case. In addition however, we will see that there are changes of polarization, in other words, amplitudes may also change as a result of the interaction.

First we fix $\theta_1$. Then $\theta_2=\theta_1+(1/\lambda_2^2-1/\lambda_1^2)t$ and so as $t\to-\infty$,
\begin{align*}
X&\sim I+{B_1\over g_1}=I+{ir_1P_1\over
e^{2\lambda_1\theta_1}-{ir_1\over 2}}.
\end{align*}
As $t\to+\infty$, using the invariance of $W$, we obtain
\begin{align*}
X&\sim I+{2B_2i\over
r_2}-{(\lambda_1+\lambda_2)(2\lambda_1B_1B_2+2\lambda_2B_2B_1-(\lambda_1+\lambda_2)r_2B_1i)-4\alpha\lambda_1\lambda_2r_1B_2i\over
r_2(\lambda_1+\lambda_2)^2g_1i-2\alpha\lambda_1\lambda_2r_1r_2}\\
&\sim
\left(I-{(\lambda_1+\lambda_2)(2\lambda_1B_1B_2+2\lambda_2B_2B_1-(\lambda_1+\lambda_2)r_2B_1i)-4\alpha\lambda_1\lambda_2r_1B_2i\over
r_2(\lambda_1+\lambda_2)^2g_1i-2\alpha\lambda_1\lambda_2r_1r_2}\left(I+{2B_2i\over
r_2}\right)\right)\left(I+{2B_2i\over
r_2}\right)\\
%&\sim I+{r_1r_2(\lambda_1+\lambda_2)^2\over
%(p_1,q_1)}{p_1q_1^t-{2\lambda_2\over
%(\lambda_1+\lambda_2)}{(p_1,q_2)\over(p_2,q_2)}p_2q_1^t-{2\lambda_2\over(\lambda_1+\lambda_2)}{(p_2,q_1)\over(p_2,q_2)}p_1q_2^t
%+{4\lambda_2^2\over(\lambda_1+\lambda_2)^2}{(p_1,q_2)(p_2,q_1)\over
%(p_2,q_2)(p_2,q_2)}p_2q_2^t\over
%r_2(\lambda_1+\lambda_2)^2(g_1+{2\alpha\lambda_1\lambda_2r_1\over(\lambda_1+\lambda_2)^2})}\\
& \sim I+{i\hat{r}_1\widehat{P}_1\over
e^{2\lambda_1\theta_1}-{i\hat{r}_1\over 2}},
\end{align*}
where
\begin{align*}
\hat{r}_1&={r_1(\hat{p}_1,\hat{q}_1)\over (p_1,q_1)},\quad \hat{P}_1={\hat{p}_1\otimes \hat{q}_1\over (\hat{p}_1,\hat{q}_1)},\\
\hat{p}_1&=p_1-{2\lambda_2\over
(\lambda_1+\lambda_2)}{(p_1,q_2)\over(p_2,q_2)}p_2,\quad
\hat{q}_1=q_1-{2\lambda_2\over
(\lambda_1+\lambda_2)}{(p_2,q_1)\over(p_2,q_2)}q_2.
\end{align*}
This shows that
\begin{align*}
W&\sim 4P_1\arctan(e^{2\lambda_1(\theta_1-\phi_1^-)}), \quad t\to -\infty,\\
W&\sim 4\widehat{P}_1\arctan(e^{2\lambda_1(\theta_1-\phi_1^+)}),
\quad t\to +\infty
\end{align*}
where
$$\phi_1^-={1\over 2\lambda_1}\log{r_1\over 2},\quad \phi_1^+={1\over 2\lambda_1}\log{r_1(\hat{p}_1,\hat{q_1})\over 2(p_1,q_1)}.$$

Similarly, fixing $\theta_2$, we have
\begin{align*}
X%&\sim I-{2B_1\over
%r_1}-{(\lambda_1+\lambda_2)(2\lambda_1B_1B_2+2\lambda_2B_2B_1-(\lambda_1+\lambda_2)r_1B_2)-4\alpha\lambda_1\lambda_2r_2B_1\over
%r_1(\lambda_1+\lambda_2)^2g_2+2\alpha\lambda_1\lambda_2r_1r_2}\\
%&\sim
%I-{(\lambda_1+\lambda_2)(2\lambda_1B_1B_2+2\lambda_2B_2B_1-(\lambda_1+\lambda_2)r_1B_2)-4\alpha\lambda_1\lambda_2r_2B_1\over
%r_1(\lambda_1+\lambda_2)^2g_2+2\alpha\lambda_1\lambda_2r_1r_2}(I-{2B_1\over
%r_1})\\
%&\sim I+{r_1r_2(\lambda_1+\lambda_2)^2\over
%(p_2,q_2)}{p_2q_2^t-{2\lambda_1\over
%(\lambda_1+\lambda_2)}{(p_1,q_2)\over(p_1,q_1)}p_2q_1^t-{2\lambda_1\over(\lambda_1+\lambda_2)}{(p_2,q_1)\over(p_1,q_1)}p_1q_2^t
%+{4\lambda_1^2\over(\lambda_1+\lambda_2)^2}{(p_1,q_2)(p_2,q_1)\over
%(p_1,q_1)(p_1,q_1)}p_1q_1^t\over
%r_1(\lambda_1+\lambda_2)^2(g_2+{2\alpha\lambda_1\lambda_2r_2\over(\lambda_1+\lambda_2)^2})}\\
%
&\sim I+{i\hat{r}_2\widehat{P}_2\over
e^{2\lambda_2\theta_2}-{i\hat{r}_2\over 2}}, \quad t\rightarrow
-\infty,\\
X&\sim I+{B_2\over g_2}=I+{ir_2P_2\over
e^{2\lambda_2\theta_2}-{ir_2\over 2}}      ,\quad t\rightarrow
+\infty,
\end{align*}
where
\begin{align*}
\hat{r}_2&={r_2(\hat{p}_2,\hat{q}_2)\over (p_2,q_2)},\quad \widehat{P}_2={\hat{p}_2\otimes\hat{q}_2\over (\hat{p}_2,\hat{q}_2)},\\
\hat{p}_2&=p_2-{2\lambda_1\over
(\lambda_1+\lambda_2)}{(p_2,q_1)\over(p_1,q_1)}p_1,\quad
\hat{q}_2=q_2-{2\lambda_1\over
(\lambda_1+\lambda_2)}{(p_1,q_2)\over(p_1,q_1)}q_1.
\end{align*}
Then
\begin{align*}
W&\sim 4\widehat{P}_2\arctan(e^{2\lambda_2(\theta_2-\phi_2^-)}),
\quad t\to
-\infty\\
W&\sim 4P_2\arctan(e^{2\lambda_2(\theta_2-\phi_2^+)}), \quad t\to
+\infty,
\end{align*}
where
$$\phi_2^-={1\over 2\lambda_2}\log{r_2(\hat{p}_2,\hat{q_2})\over 2(p_2,q_2)}, \quad \phi_2^+={1\over 2\lambda_2}\log{r_2\over 2}.$$

The above calculations show that $W(x,t)$ decomposes into the sum of two kink
solutions as $t\to\pm\infty$ and the $j$th kink solution propagates
with the speed $1/\lambda_j^2$. The phase shifts
$\Delta_j=\phi_j^+-\phi_j^-$ for the kink solutions  are
\[
\Delta_1={1\over 2\lambda_1}\log\beta, \quad \Delta_2=-{1\over
2\lambda_2}\log\beta,
\]
where
\[
\beta=1-{4\lambda_1\lambda_2\alpha\over (\lambda_1+\lambda_2)^2}.
\]
\begin{rem}
In a similar way to the matrix KdV equation in \cite{VMG}, we find that the matrix
amplitude of the first kink solution changes from $4P_1$ to
$4\widehat{P}_1$ and the matrix amplitude of the other one changes from
$4\widehat{P}_2$ to $4P_2$ as $t$ changes from $-\infty$ to $+\infty$.
If $(p_1,q_2)=0$ ($P_2P_1=0$) or $(p_2,q_1)=0$ ($P_1P_2=0$), there is no phase-shift, however
the amplitudes may change. In the case that both $P_1P_2=P_2P_1=0$, there is neither phase-shift nor change in amplitude and so the kink solutions have trivial interaction.
\end{rem}

\begin{figure}\centering
    \begin{tabular}{cc}
        \includegraphics*[width=3in]{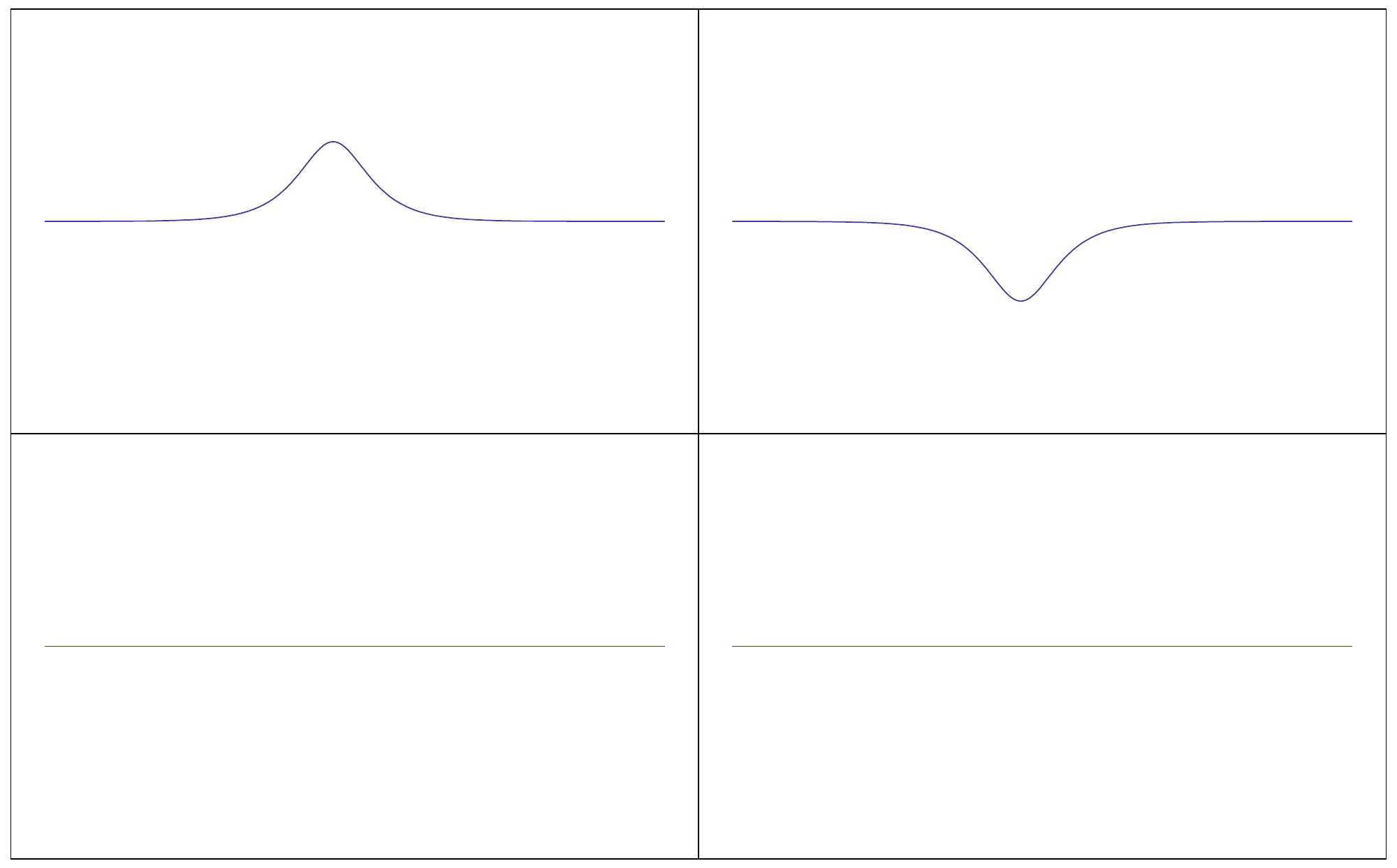}&\includegraphics*[width=3in]{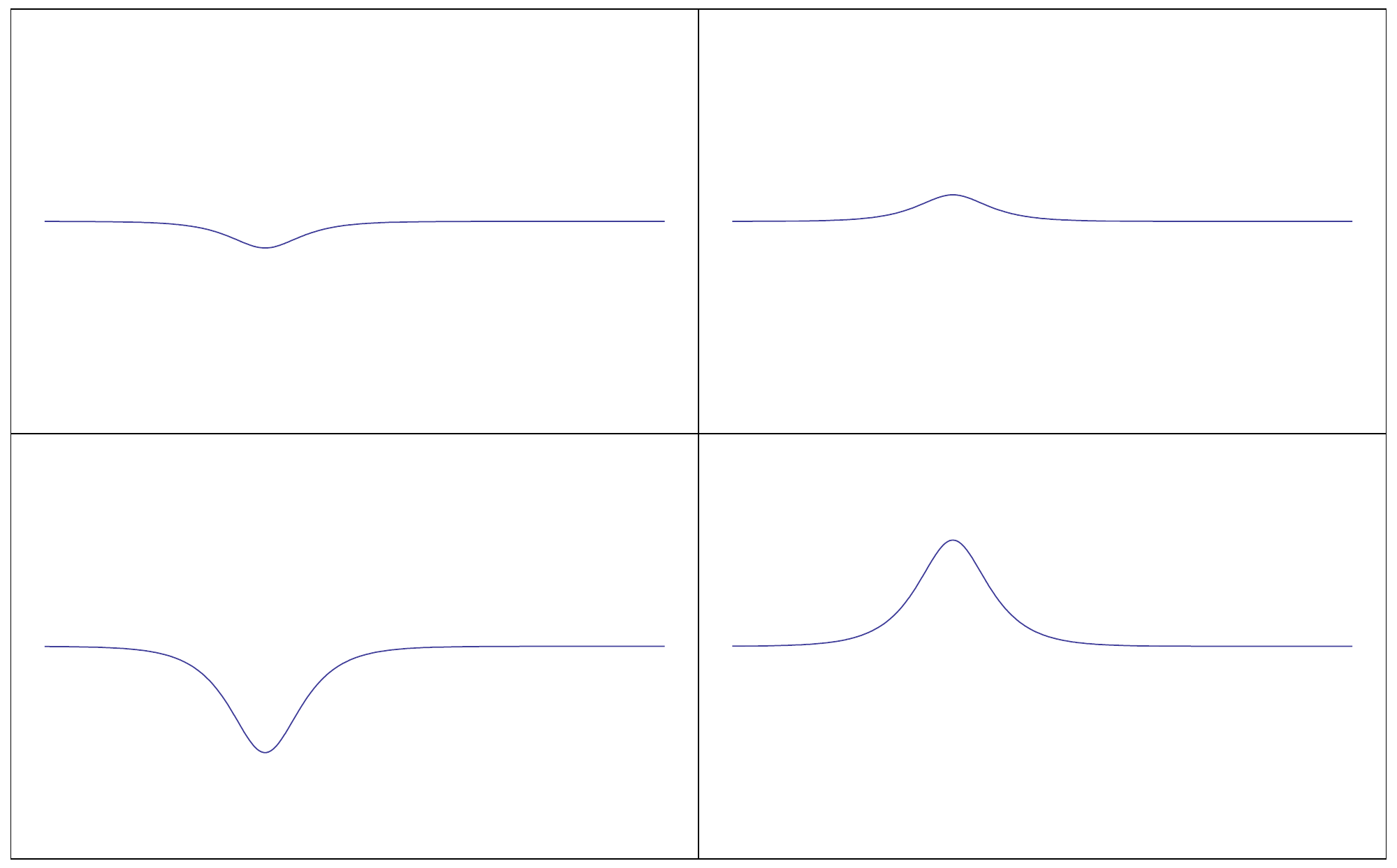} \\
    $\displaystyle P_1^{-}=\left(\begin{array}{cc}1&-1\\0&0\end{array}\right)$&$\displaystyle P_1^{+}={1\over 3}\left(\begin{array}{cc}-1&1\\-4&4\end{array}\right)$
    \end{tabular}
   \caption{
Asymptotic forms for kink 1
\label{fig1}
}
\end{figure}

\begin{figure}\centering
    \begin{tabular}{cc}
        \includegraphics*[width=3in]{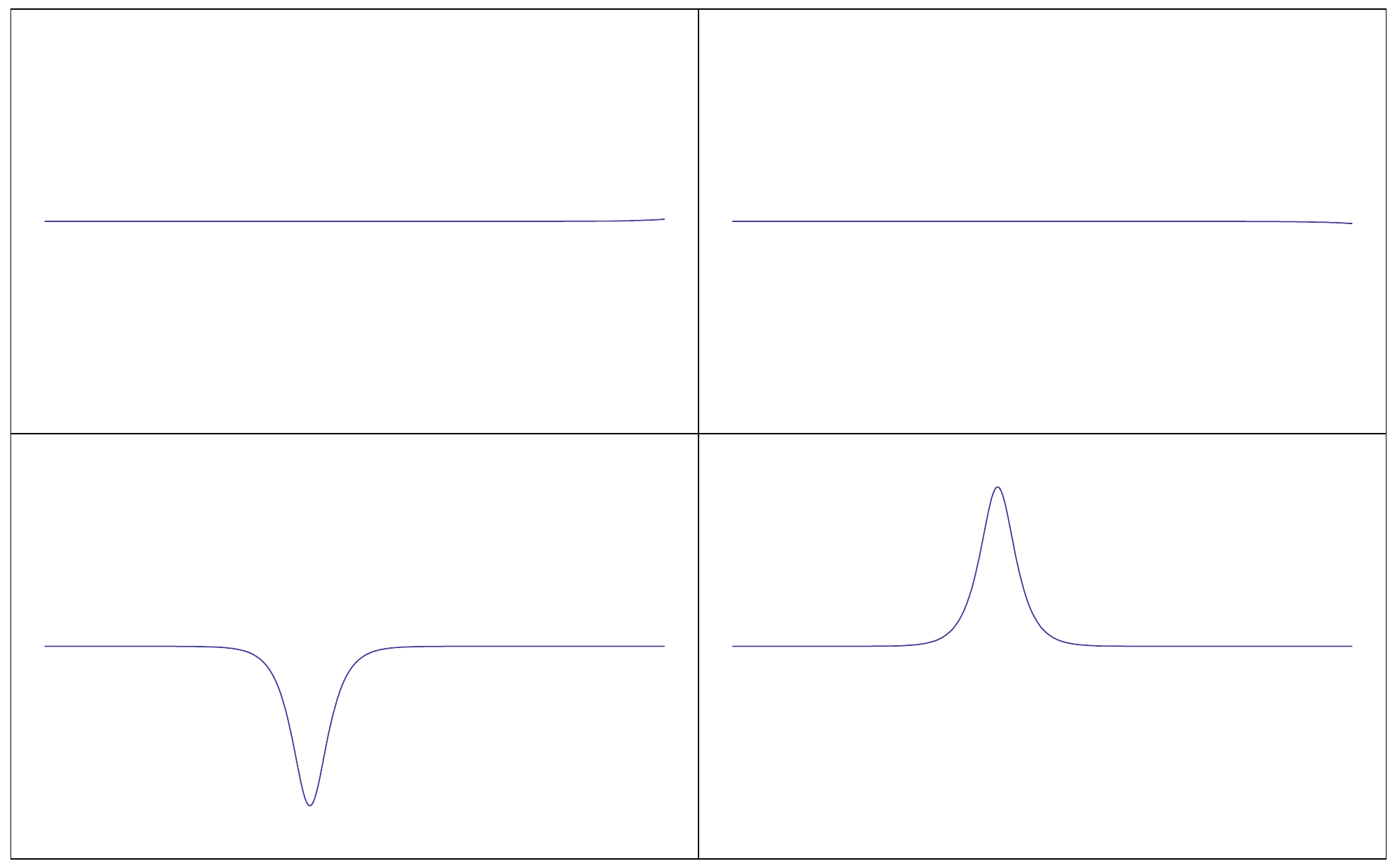}&\includegraphics*[width=3in]{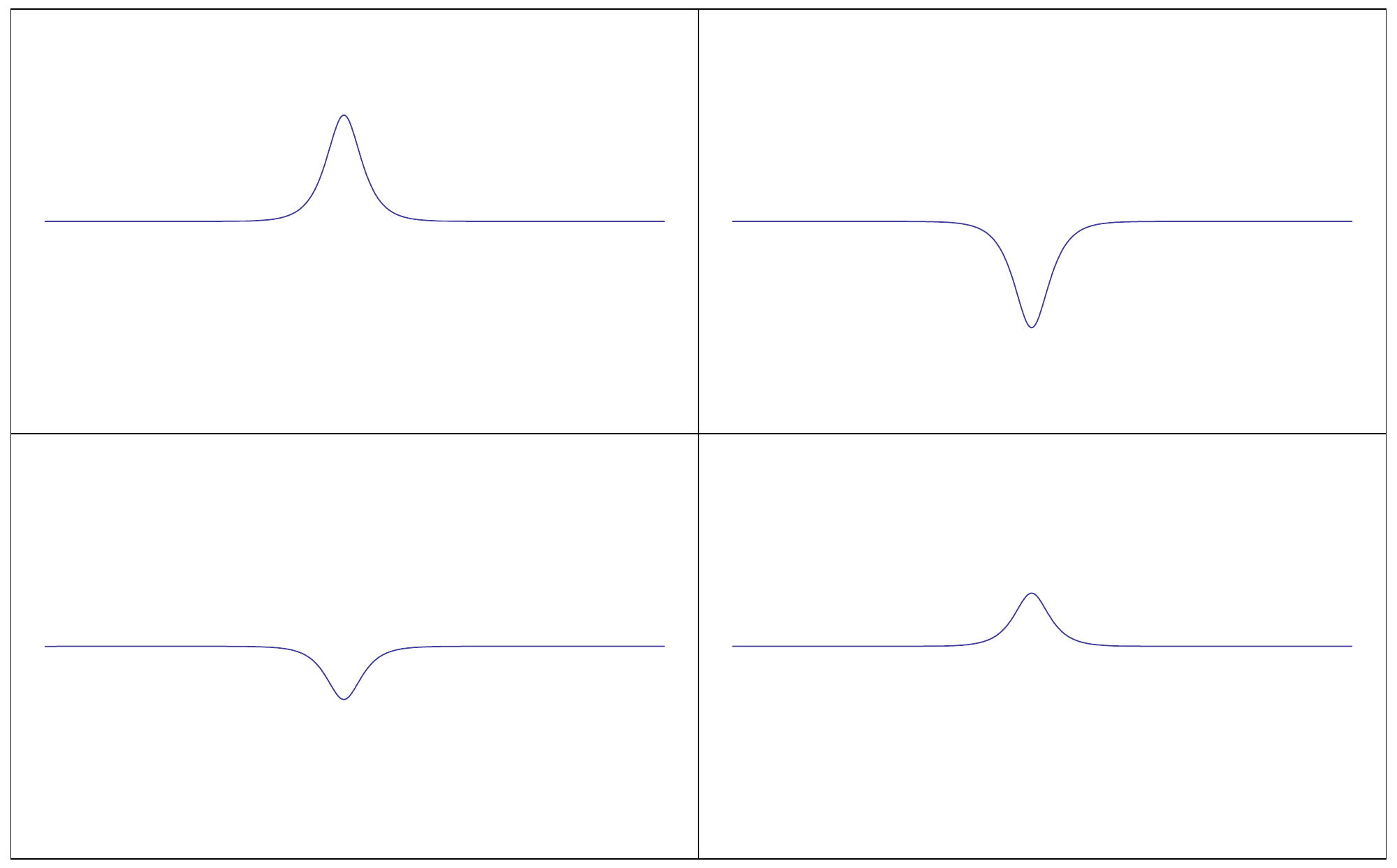} \\
    $\displaystyle P_2^{-}=\left(\begin{array}{cc}0&0\\-1&1\end{array}\right)$&$\displaystyle P_2^{+}={1\over 3}\left(\begin{array}{cc}2&-2\\-1&1\end{array}\right)$
    \end{tabular}
\caption{
Asymptotic forms for kink 2
\label{fig2}
}
\end{figure}

To illustrate the above, we will consider the case $d=2$, i.e. the $2\times 2$ matrix sine-Gordon equation. We choose $\lambda_1=1,\ \lambda_2=2,\ r_1=r_2=1$, and
\[
    P_1=\left(\begin{array}{cc}1&-1\\0&0\end{array}\right),\quad
    P_2={1\over 3}\left(\begin{array}{cc}2&-2\\-1&1\end{array}\right).
\]
The analysis above shows that $P_1^-=P_1$, $P_2^+=P_2$ and
\[
    P_1^{+}={1\over 3}\left(\begin{array}{cc}-1&1\\-4&4\end{array}\right),\quad
    P_2^{-}=\left(\begin{array}{cc}0&0\\-1&1\end{array}\right).
\]
For convenience, rather than plotting the kink $W$ given by \eqref{soliton}, we plot the derivative $W_x$ and refer to it as a soliton. In Figure~\ref{fig1}, the asymptotic forms of the matrix soliton 1 are plotted as $t\to\pm\infty$. The first plot exhibits the amplitudes given by $P_1^-$ and the second, those of $P_1^+$. Similarly, in Figure~\ref{fig2}, we show the same plots for soliton 2.

\section{Conclusions}
In this paper, we have considered a generalized non-Abelian Toda lattice and presented quasiwronskian and quasigrammian solutions obtained means of by Darboux transformations and binary Darboux transformations respectively. Then we imposed a $2$-periodic reduction on the non-Abelian Toda lattice to derive a noncommutative sine-Gordon equation. By using a method similar to that developed in \cite{VMG} for the matrix KdV equation, we obtained kink solutions for the matrix sine-Gordon equation from the quasigrammian solutions of the non-Abelian Toda lattice. Then we investigated the interaction properties of matrix kink solutions. It is known \cite{VAP} that the change of matrix amplitude of solitons for the matrix KdV equation gives rise to a Yang-Baxter map. It would be interesting to investigate whether there is a similar result for the matrix sine-Gordon equation.

\section*{Acknowledgement}
This work was supported by the Royal Society China Fellowship and
the National Natural Science Foundation of China (grant No.
10601028). We would also like to thank the referees for their valuable comments.

\bibliography{refs}

\end{document}